\documentclass{revtex4}
\usepackage{graphicx}
\usepackage{fancyhdr}
\usepackage{amsmath}
\newcommand{\bea}{\begin{eqnarray}}
\newcommand{\eea}{\end{eqnarray}}

\usepackage{color}

\begin{document}

\title{Higgs Boson Mass in Low Energy SUSY Models with Vector-like Matters}

%

\author{Norimi Yokozaki}
\affiliation{Kavli IPMU, University of Tokyo, Kashiwa, Chiba 277-8568, JAPAN}

\begin{abstract}
We show that the gauge mediation models with vector-like matters can explain the Higgs boson mass of 126 GeV without large soft scalar masses nor a large left-right mixing of the stops. The scenario has interesting features: the existence of the light non-colored SUSY particles accessible at the LHC and ILC, the explanation of the muon $g-2$ and the possibility of the enhanced di-photon decay rate of the Higgs boson.

\end{abstract}

\maketitle

\thispagestyle{fancy}


\section{Introduction}
The discovery of the Higgs boson like particle with the mass of $\sim\,$126\,GeV has great influence on the supersymmetric models. In the minimal supersymmetric standard model (MSSM), the Higgs boson mass at the tree level is at most $\sim91$\,GeV, and hence, the Higgs boson mass of $\sim\,$126\,GeV requires large radiative corrections; the quartic coupling in the Higgs potential should be twice as large as the tree level value. The required corrections can be obtained by heavy stops of $\mathcal{O}(10-1000)$\,TeV or a large left-right mixing for $\mathcal{O}(1)$\,TeV stops~\cite{SUSYhiggs}. In the former case, the SUSY CP/flavor problem as well as the cosmological moduli problem is significantly relaxed, while not in the latter case \footnote{For instance, the gaugino mediation models with low scale cutoff can solve to these problems~\cite{gaugino_med}.}.
 Therefore, the heavy SUSY scenarios are somewhat favored, combined with the non-observation of the SUSY signals at the LHC~\cite{SUSY_ATLAS, SUSY_CMS}.

However, there is an indication that the SUSY particles, at least, non-colored SUSY particles are light as $\mathcal{O}(100)$ GeV: the anomalous magnetic moment of the muon (muon $g-2$). The experimental value of the muon $g-2$~\cite{Bennett:2006fi} is deviated from the standard model (SM) prediction~\cite{Davier:2010nc, Hagiwara:2011af} at more than $3\,\sigma$ level. Using the SM prediction in Ref.~\cite{Hagiwara:2011af}, the deviation is given by
\bea
\delta a_\mu \equiv a_\mu^{\rm EXP}-a_\mu^{\rm SM}=(26.1\pm8.1)\times10^{-10}.
\eea
It was pointed out that, in MSSM, $\delta a_\mu$ can be explained by the contributions from the light smuons and chargino/neutralino with the mass of $\mathcal{O}(100)$\,GeV for $\tan\beta=\mathcal{O}(10)$~\cite{SUSY_gm2}. If the deviation is true, there exists a tension; the Higgs boson mass favors heavy SUSY particles while the muon $g-2$ requires light SUSY particles with $\mathcal{O}(10)\,\tan\beta$. Because of this, there are not many SUSY scenarios which can accommodate both of them. For instance, SUSY models with a singlet extension of the Higgs sector can not explain the muon $g-2$, since the enhancement of the Higgs boson mass occurs only in the region with low $\tan\beta$. 

In this letter, we consider the gauge mediation models with the vector-like matters~\cite{VGMSB} as one of the few scenarios consistent with the muon $g-2$ experiment and the Higgs boson mass of around 126\,GeV~\footnote{Other scenarios can be found in Refs.~\cite{other1, other2, other3, other4}}. The   vector-like matters with $\sim 1$\,TeV mass which couple to the Higgs boson chiral multiplet can enhance the Higgs boson mass by the radiative corrections in a similar manner of the top/stop loops~\cite{OM, babu, martin}, even for large $\tan\beta$. Because of this effect, the Higgs boson mass is explained with relatively light colored SUSY particles of $1-2$ TeV, allowing the existence of the light non-colored SUSY particles of $\mathcal{O}(100)$\,GeV; the scenario gives rich phenomenologies, such as the muon $g-2$ explanation and  the possibility of the enhanced di-photon decay rate of the Higgs boson.

\section{Gauge Mediation Models with Vector-like Matters}
We consider the gauge mediation models with extra vector-like matters added to the MSSM matter contents. The vector-like matters are introduced as ${\bf 10}$ and ${\bf \overline{10}}$ representation in SU(5) grand unified theory (GUT) gauge group so that the gauge coupling unification is maintained. Then, the superpotential of the matter sector is given by
\bea
W=W_{\rm MSSM-Yukawa} + Y' Q' H_u \overline{T'} + \mu H_u H_d + M_Q Q'\overline{Q'} + M_T T'\overline{T'} + M_E E'\overline{E'} ,
\eea
where ${\bf10}=(Q', \overline{T'}, \overline{E'})$ ( and ${\bf \overline{10}}$ ) are additional vector-like matters. The SUSY invariant mass terms, $\mu, M_Q, M_T$ and $M_E$, are possibly related to the Peccei-Quinn (PQ) symmetry breaking scale~\cite{nakayama_yokozaki}. Here, we neglect $\overline{Q'} H_d U'$, which can be suppressed by the PQ symmetry. There are new radiative corrections through the term $Y' Q' H_u \overline{T'}$, which are similar to the top/stop loops. With a large Yukawa coupling $Y' \simeq 1$ and $M_Q\sim M_T\sim 1$\,TeV, the Higgs boson mass can be efficiently enhanced and can reach to $\sim126$\,GeV without the large left-right mixing nor heavy scalar masses. The leading part of the radiative correction to the Higgs boson mass from the extra-matters is 
\bea
\Delta m_h^2 \sim \frac{3 Y'^4 v^2}{4\pi^2} \sin^4\beta \left[ \ln \frac{m_{\rm squark}^2}{M_F^2} \right], 
\eea
where $m_{\rm squark}$ is the soft mass of the extra squark and $M_F \sim M_Q \sim M_U$. The vacuum expectation value of the Higgs boson is denoted by $v\, (\simeq 174$\,GeV). 
For the  hierarchical SUSY mass and SUSY breaking mass generate sufficiently large $\Delta m_h^2$. 
Note that $Y' \simeq 1$ at around the weak scale is quite natural since it has a quasi infrared fix point with the value of unity~\cite{martin}.

The soft SUSY breaking mass parameters are induced by messenger loops. The superpotential of the messenger sector is written as
\bea
W_{\rm mess}=(M_D + F_D \theta^2) \Psi_D \Psi_{\bar{D}} + (M_L + F_L \theta^2) \Psi_L \Psi_{\bar{L}} ,
\eea
where $\Psi_L$ and $\Psi_{\bar{D}}$ are $SU(2)_L$ doublet and $SU(3)_C$ triplet messengers, respectively. The hyper-charge of $\Psi_L$ ($\Psi_{\bar{D}}$) is $-1/2$ ($1/3$). If relations, $M_D=M_L$ and $F_D=F_L$ are hold at the GUT scale, $F_D/M_D \simeq F_L/M_L$ is hold at any scale. However, such relations can be violated by, e.g., a higher dimensional operator picking up the GUT scale VEV. Here, we consider only a pair of the messengers. This is because introducing two or more messengers leads to the Landau pole below the GUT scale, unless the messenger scale is sufficiently high.

The leading contributions to gaugino masses and scalar masses at the messenger scale are given by
\bea
M_1\simeq \frac{g_1^2}{16\pi^2} \left[ (2/5)  (F_D/M_D) + (3/5)  (F_L/M_L) \right], \ M_2\simeq \frac{g_2^2}{16\pi^2} (F_L/M_L), \ M_3\simeq \frac{g_3^2}{16\pi^2}  (F_D/M_D),
\eea
and
\bea
m_Q^2 &\simeq& \frac{2}{(16\pi^2)^2} \left[ g_3^4 (4/3) (F_D/M_D)^2 +g_2^4 (3/4) (F_L/M_L)^2 \right.\nonumber \\
&+& \left. g_1^4 (1/60) ( (2/5)(F_D/M_D)^2 + (3/5)(F_L/M_L)^2 )\right],\nonumber \\
m_U^2 &\simeq& \frac{2}{(16\pi^2)^2} \left[g_3^4 (4/3) (F_D/M_D)^2 \right. \nonumber \\ 
&+& \left. g_1^4 (4/15) ( (2/5)(F_D/M_D)^2 + (3/5)(F_L/M_L)^2 )\right],\nonumber \\
m_D^2 &\simeq& \frac{2}{(16\pi^2)^2} \left[g_3^4 (4/3) (F_D/M_D)^2 \right. \nonumber \\
&+& \left. g_1^4 (1/15) ( (2/5)(F_D/M_D)^2 + (3/5)(F_L/M_L)^2 )\right] ,\nonumber \\
m_L^2 &\simeq& \frac{2}{(16\pi^2)^2} \left[g_2^4 (3/4) (F_L/M_L)^2 \right. \nonumber \\
&+& \left. g_1^4 (3/20) ( (2/5)(F_D/M_D)^2 + (3/5)(F_L/M_L)^2 )\right],\nonumber \\
m_E^2 &\simeq& \frac{2}{(16\pi^2)^2} \left[g_1^4 (3/5)( (2/5)(F_D/M_D)^2 + (3/5)(F_L/M_L)^2 )\right],\nonumber \\
m_{H_u}^2 &=& m_{H_d}^2=m_L^2,
\eea
where $M_1$, $M_2$ and $M_3$ are the bino, wino and gluino, respectively. The mass of the SU(2) doublet squark is denoted by $m_{Q}$ while those of the $SU(2)$ singlet are denoted by $m_{U}$ (up-type) and $m_D$ (down-type). The slepton masses are written as $m_{L}$ and $m_E$ for $SU(2)$ doublet and singlet, respectively. By taking the ratio $(F_L/M_L)/(F_D/M_D)$ to be smaller than unity, we can split the masses of the colored and non-colored SUSY particles, which enhances the SUSY contributions to the muon $g-2$ for the fixed masses of the colored SUSY particles.

\begin{figure}[h]
\centering
\includegraphics[width=120mm]{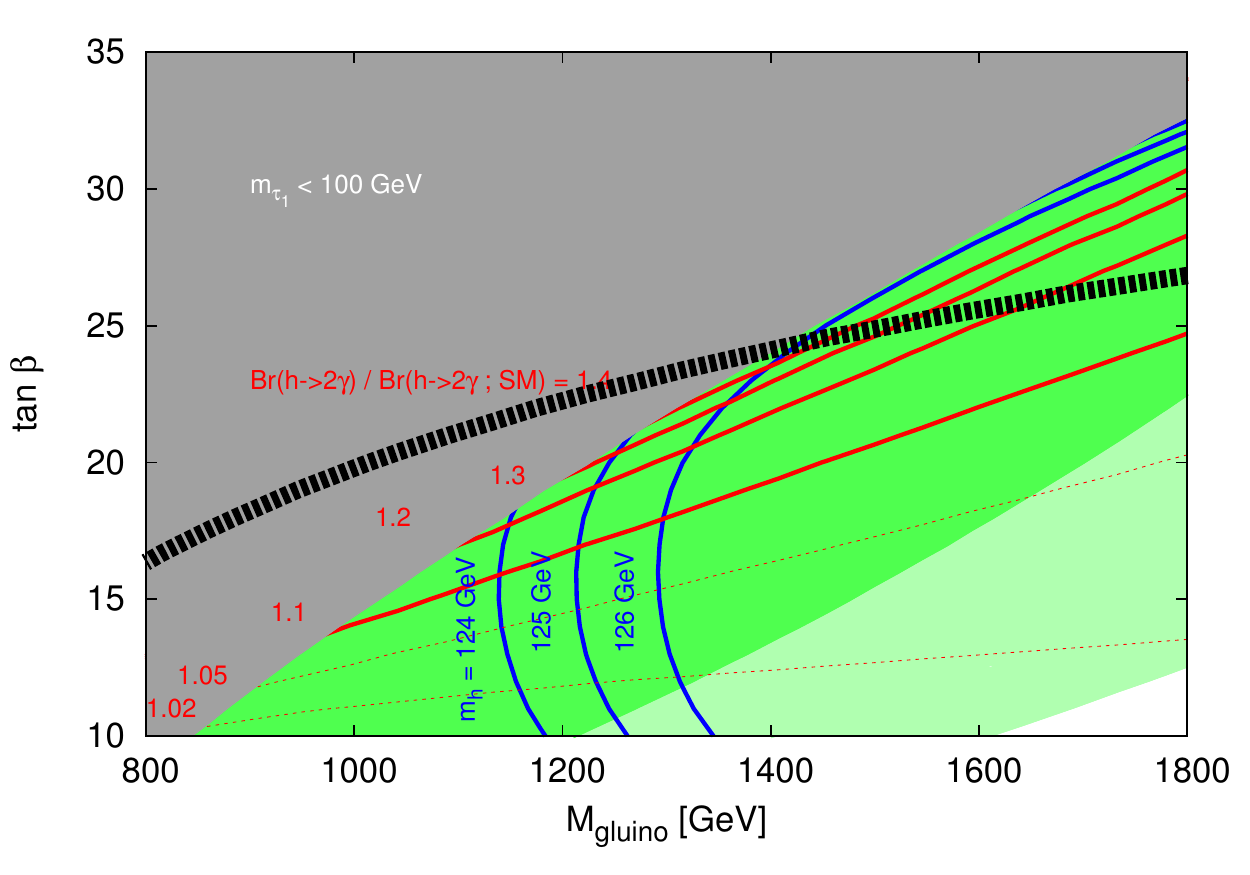}
\caption{Contours of the Higgs boson mass (blue solid line) and the region consistent with the muon $g-2$. The Higgs to di-photon ratio $r_{\gamma \gamma}$ is also shown (red solid line). The Higgs boson mass is calculated with $M_Q=M_T=900$\,GeV, $m_{t}({\rm pole})=173.3$ GeV and $Y'(m_{\rm SUSY})=1.0$. In the deep (light) green region, the muon $g-2$ is explained at 1$\sigma$ (2$\sigma$) level. The black dashed line shows the stability bound of the electroweak symmetry breaking (EWSB) minimum. The region above the line is excluded since the lifetime of the EWSB minimum becomes shorter than the age of the universe. Here, we take the messenger scale and SUSY breaking masses as $M_{D}=M_{L}=4\times10^5$ GeV and $(F_L/F_D)=0.6$, respectively.
}\label{fig:vector}
\end{figure}

The contours of the Higgs boson mass and the region consistent with the muon $g-2$ are shown on the gluino mass - $\tan\beta$ plane in Fig.\,\ref{fig:vector}. In the deep (light) green region, $\delta a_\mu$ is explained by the SUSY contributions at 1$\sigma$ (2$\sigma$) level. The Higgs boson mass is calculated with $M_Q=M_T=900$\,GeV. Taking smaller value of $M_Q$ (and $M_T$) makes the Higgs boson mass larger while larger $M_Q$ makes the mass smaller. The region above the black dashed line is excluded by the vacuum stability condition; there exists a charge breaking global minimum induced by the large left-right mixing of the stau ($\sim m_\tau \mu \tan\beta$) and the electroweak symmetry breaking (EWSB) minimum is unstable. Requiring that the life-time of the EWSB minimum be longer than the age of the universe, $\mu\tan\beta$ can not be too large~\cite{stau_vacuum}. 

The contours of the ratio of the Higgs branching fraction to di-photon, $r_{\gamma\gamma}$, are also shown. Here, $r_{\gamma \gamma}$ is defined as $r_{\gamma \gamma} \equiv {\rm Br}(h \to \gamma\gamma)/{\rm Br}(h_{\rm SM} \to \gamma\gamma)$. It was shown that the di-photon decay rate is enhanced by the stau loops with the large left-right mixing~\cite{diphoton}. In our case, the di-photon ratio can be enhanced up to about 40\% compared to the SM prediction. The enhancement is bounded from above by the vacuum stability constraint~\cite{diphoton_vac}.

In most of the region consistent with the muon $g-2$ at 1$\sigma$ level, the stau is the next-to-lightest SUSY particle (NLSP) with the mass less than about 200\,GeV. The light (quasi-)stable stau is severely constrained from the LHC. The stau mass should be larger than about 340\,GeV~\cite{CMS_stau}. On the other hand, the stau can decay with a relatively short decay length, say, less than $\sim$10\,cm. For exmaple, the stau decays into the tau plus gravitino or axino. The former case corresponds to the ultra-light gravitino with the mass of less than about 10\,eV,  the latter case is realized in the DFSZ axion model~\cite{DFSZ} with the small axion decay constant $f_a \sim 10^9$ GeV. Although the production of the stau through the colored SUSY particles can be suppressed, the production through the chargino/neutralino is not; the chargino mass should be larger than about 280 GeV~\cite{CMS_stau_decay}. The sample mass spectrum is shown in Table.\,1.


%



\begin{table}[h]
\begin{center}
\caption{A sample mass spectrum}
\begin{tabular}{l|c}
$M_{\rm mess}$ & 300\,TeV  \\
$F_D$ & 170\,TeV \\
$F_L/F_D$ & 0.6  \\
$\tan\beta$ & 20  \\
\hline
\hline $m_{\rm gluino}$ & 1.6\,TeV \\
\hline $m_{\rm squark}$ & 2.4\,TeV  \\
\hline $m_{L}$ & 413\,GeV   \\
\hline $m_{E}$ & 243\,GeV \\
\hline $m_{\tilde{\tau}_1}$ & 165\,GeV \\
\hline $m_{\chi_1^0}$ & 190\,GeV \\
\hline $m_{\chi_1^\pm}$ & 302\,GeV \\
\hline $\delta a_\mu$ & $21.4 \times 10^{-10}$
\end{tabular}
\label{example_table}
\end{center}
\end{table}

\section{Conclusion}
In this letter, we have considered a gauge mediation model with the vector-like matters at around 1\,TeV. With the help of the radiative corrections from the vector-like matters, this scenario can explain the Higgs boson mass of $\sim 126$\,GeV without large scalar masses nor the large left-right mixing, which is suitable for the gauge mediated SUSY breaking scenarios. Because of the light non-colored SUSY particles,  the muon $g-2$ deviation is successfully explained. Interestingly, in some region, the di-photon decay rate of the Higgs boson can be enhanced up to 40\%. It has been shown that the constraint from the LHC SUSY search can be avoided, but still, the (right handed) sleptons are lighter than $\sim250$ GeV and can be target at the International Linear Collider experiment.





\begin{acknowledgments}
N.Y. thanks organizers of {\it ``Toyama International Workshop on Higgs as a Probe of New Physics''} for hospitality.
\end{acknowledgments}

\bigskip 

\end{document}